\documentclass[times, 10pt,twocolumn]{article} 

\usepackage{latex8}
\usepackage{times}

\usepackage{graphicx}
\usepackage[figuresright]{rotating}

\begin{document}

\title{3D Computer Simulations of Pulsatile Human Blood
Flows in Vessels and in the Aortic Arch: Investigation of Non-Newtonian
Characteristics of Human Blood}

\author{Renat A. Sultanov\\
Business Computing Research Laboratory, St. Cloud State University\\
Centennial Hall, 4-th Avenue South, 367 B\\
St. Cloud, MN 56301, USA\\ rasultanov@stcloudstate.edu\\ 
\and
Dennis Guster\\
Business Computer Information Systems, St. Cloud State University\\
Centennial Hall, 4-th Avenue South, 367 C\\
St. Cloud, MN 56301, USA \\dcguster@stcloudstate.edu\\ 
\and
Brent Engelbrekt$^{(1)}$ and Richard Blankenbecler$^{(2)}$\\
RIE Coatings/RIE Medical LLC., 221 Logeais Street, P.O. Box 350\\
Eden Valley, MN 55329, USA \\brent@meyercommercial.com$^{(1)}$; rblankenbecler@cox.net$^{(2)}$}

\maketitle
\thispagestyle{empty}

\begin{abstract}
Methods of Computational Fluid Dynamics are applied to simulate
pulsatile blood flow in human vessels and in the aortic arch. The 
non-Newtonian behaviour of
the human blood is investigated in simple vessels of actual size.
A detailed time-dependent mathematical convergence test has been carried out.
The realistic pulsatile flow is used in all simulations. Results of
computer simulations of the blood flow in vessels of two different geometries are
presented. For pressure, strain rate and velocity component
distributions we found significant disagreements between our results obtained 
with realistic non-Newtonian treatment of human blood and widely
used method in literature: a simple Newtonian approximation. A significant
increase of the strain rate and, as a result,
wall sear stress distribution, is found in the
region of the aortic arch. We consider this result as
theoretical evidence that supports existing
clinical observations and those models not using non-Newtonian treatment
underestimate the risk of disruption to the human vascular system.
\end{abstract}



\Section{Introduction}

Human blood is a liquid with variable density and viscosity.
The movement of the blood inside vessels and arteries can be described by
fundamental laws of physics, i.e. equations of fluid dynamics. The
scientific literature now contains many citations where researchers
have used computer simulations of blood flows in various size vessels
and arteries at different spatial geometries, see for example [1-10].
Whereas experimental investigations of vascular dynamics and flow 
are complicated or simply impossible to carry out due to small sizes
of vessels in living systems. Therefore, theoretical-mathematical models and
computer simulations are very useful for studying blood flows.

Atherosclerosis occurs at specific arterial sites.
This phenomenon is related to hemodynamics and to wall shear stress
(WSS) distributions.
WSS is the tangential drag force produced by moving blood.
It is a mathematical function of the velocity gradient of blood near
the endothelial surface:
$
%
\tau_w=\mu \left [\partial U(t,y,R_v) / \partial y \right ]_{y \approx 0,}
%
$
here $\mu$ is the dynamic viscosity, $t$ is current time, $U(t,y,R_v)$
is the flow velocity parallel to the wall, $y$ is the distance to
the wall of the vessel, and $R_v$ is its radius.
It was shown, that the magnitude of WSS
is directly proportional to blood flow and blood viscosity and
inversely proportional to the cube of the radius of vessels,
in other words a small change of the radius of a vessel will have a large 
effect on WSS.

Arterial wall remodeling is regulated by WSS. In response to high
shear stress arteries enlarge. Consequently, 
the atherosclerotic plaques localize preferentially in regions of
low shear stresses, but not in regions of higher shear stresses.
Furthermore, decreased shear stress induces intimal thickening in vessels which
have adapted to high flow.
Also, final vascular events that induce fatal outcomes, such as 
acute coronary syndrome, are triggered by the sudden mechanical disruption of an arterial 
wall. Thus, we can conclude, that the final consequences of tragic fatal vascular diseases 
are strongly connected to mechanical events that occur on the vascular
wall, and these, in turn, which are likely to be heavily influenced by
alterations in blood flow and the characteristics of the blood itself.


In order to predict, diagnose, and prevent fatal outcomes in these vascular diseases, 
{\it fluid + solid mechanical} interactions between the human blood and the vascular wall 
are attractive and necessary targets for analysis. However, it is very
difficult to make measurements of detailed mechanical properties in
living systems. In turn, theoretical bio-mathematics provides a series of logic
tools that can overcome limitations of direct measurements in highly
labile living systems and provide a framework for testing variables,
and is more rapid, efficient, and possibly more predictive than
repeated experimentation in animal modeling systems

In this work we carry out real-time full-dimensional
computer simulations of pulsatile blood
flows in actual size vessels and in the aortic arch.
We take into account different physical effects on blood flow
and examine the non-Newtonian nature of human blood as a
fluid. 

Computer simulation is
well-suited to those cases in which it is difficult to carry out
reliable experiments due to the very small size of some vessels, such
as, for example, coronary arteries.
Also, 
bio-mathematical computer simulations may be especially useful in
situations when an intravascular stent is implanted inside a
vessel, aortic arch, aortas etcetera. Stents may have complicated shapes and they are very
small micro-devices (typically only several millimeters in diameter, 
when fully deployed). It is useful to know pressure, strain rate
distributions and the profile of blood flow after coming through a stented vessel.


The next section presents the mathematical and numerical methods used in this
work. Section 3 presents our results. 
The {\sf CGS} unit system is used in
all simulations as well as for presentation of our results.

\Section{Method}

In this work we consider the human blood as an incompressible fluid, and
flows in vessels and in the aortic arch are assumed to be laminar.
One of the goals of this work is to investigate the non-Newtonian
behaviour of the human blood. To attain these ends we carry out
simulations for same systems, but with different models of the blood. Then
we compare the results.

For many years, investigation of hemorheology has been of great interest in the field of 
biomedical engineering. Researchers have investigated correlations for
example, of stroke, arterial diseases, hypertension, and whole blood 
viscosity. 
Blood consists of plasma and particles, including red blood cells,
leukocytes, platelets and macromolecular protein aggregates. The
viscosity of  blood depends on the viscosity of plasma, in combination
with the hematocrit (a measure of the particulate component of
blood). The normal hematocrit of human blood ranges between 35\% - 45\%.
Higher hematocrit implies higher viscosity. 
The relation between hematocrit and viscosity is very complex and in the scientific
literature, many mathematical fitting formulas are available for
assessing this relationship. 

Next, the viscosity of blood determines its velocity. That is, when velocity
or shear rate increases viscosity decreases.
Also, the viscosity/velocity depends on the 
size of the blood vessel. This is called the Fahraeus-Lindqvist
effect, 
that is in small diameter blood vessels, and at higher velocities, blood
viscosity decreases. Viscosity of human blood strongly depends on its temperature. 


To carry out our simulations we used a commercial program FLOW3D from Flow
Science Inc., Santa Fe, New Mexico, USA.
FLOW3D is a general purpose CFD package. It applies specially
developed numerical techniques to solve the equations of motion of
fluid. The methods implemented in the program allow us to obtain
transient, 3D solutions for multi-scale and multi-physics flow
problems. One of the most attarctive sides of FLOW3D is very well
developed numerical techniques to solve non-linear fluid dynamic equations.
To our best knowledge our work is the first time application of the FLOW3D
program to blood flows in human vessels.

When the turbulence option is used, the viscosity is a sum of the
molecular and turbulent values. For non-Newtonian 
fluids the viscosity can be a function of the 
strain rate and/or temperature. A general expression
based on the Carreau model is used in FLOW-3D
for the strain rate dependent viscosity:
\begin{eqnarray}
\mu = \mu_{\infty}+\frac{\mu_0E_T-\mu_{\infty}}{\lambda_{00}+
[\lambda_0+(\lambda_1E_T)^2e_{ij}e_{ij}]^{(1-n)/2}}\nonumber \\
+ \frac{\lambda_2}{\surd (e_{ij}e_{ij})},
\label{eq:mu1}
\end{eqnarray}
where
%
$
e_{ij}=1/2(\partial u_i/\partial x_j +\partial u_j/\partial x_i)
$
%
is the fluid strain rate in Cartesian tensor notations, $\mu_{\infty},
\mu_0, \lambda_0, \lambda_1, \lambda_2$ and $n$ are constants. Also,
$
E_T = exp[a(T^*/(T-b) - C)],
$
%
where $T^*, a, b,$ and $c$ are also parameters of the temperature
dependence, and $T$ is fluid temperature. This basic formula is used
in our simulations for blood flow in vessels and in the aortic arch.

Next, fluid dynamics is described with 2-nd order non-linear, transient
differential equations. The governing equations consist of the continuity
equation and the Navier-Stokes equations. The general mass continuity
equation, which is solved within the FLOW3D program has the following
{\it general} form:
\begin{eqnarray}
V_f\frac{\partial \rho}{\partial t} + 
\frac{\partial}{\partial x}(\rho u A_x)+
R\frac{\partial}{\partial y}(\rho v A_y)+
\frac{\partial}{\partial z}(\rho w A_z)+\nonumber \\
\xi \frac{\rho u A_x}{x} = R_{dif}+R_{sor},
\end{eqnarray}
where $V_F$ is the fractional volume open to flow, 
$R$ and $xi$ are coefficients which value dependent on the coordinate system:
$(x,y,z)$ or $(r,\theta,z)$, $\rho$ is the fluid density,
$R_{dif}$ is a turbulent diffusion term, and $R_{sor}$ is a mass
source, $(u, v, w)$ are the velocity components in coordinate
directions $(x,y,z)$ respectively. For example,
when Cartesian coordinates are used, $R=1$ and $\xi=0$, 
see FLOW3D manual \cite{flow3d}. Finally, $A_x$ is the fractional area
open to flow in the $x$ direction, analogously for $A_y$ and $A_z$.

The turbulent diffusion term is
\begin{eqnarray}
R_{dif}=
\frac{\partial}{\partial x}(v_pA_x\frac{\partial \rho}{\partial x})+R
\frac{\partial}{\partial y}(v_pA_yR\frac{\partial \rho}{\partial
  y})+\nonumber \\
\frac{\partial}{\partial z}(v_pA_z\frac{\partial \rho}{\partial z})+
\xi\frac{\rho v_pA_x}{x},
\end{eqnarray}
where the coefficient $v_p=C_p\mu / \rho$, $\mu$ is dynamic viscosity and
$C_p$ is a constant. The $R_{sor}$ term is a density source term that
can be used to model mass injections through porous obstacle surfaces.

It is well known, that compressible flow problems require solution of the
full density transport equation. In this work we treat blood as
an incompressible fluid. For incompressible fluids $\rho=constant$ and
the Eqn. (2) becomes the following:
\begin{equation}
\frac{\partial}{\partial x}(u A_x)+
\frac{\partial}{\partial y}(v A_y)+
\frac{\partial}{\partial z}(w A_z)+
\xi \frac{u A_x}{x} = \frac{R_{sor}}{\rho}.
\end{equation}

The equations of motion for the fluid velocity components $(u, v, w)$
in the 3-coordinate system are the Navier-Stokes equations
with specific additional terms included in the FLOW3D program:
\begin{eqnarray}
\frac{\partial u}{\partial t}+\frac{1}{V_F}
{uA_x\frac{\partial u}{\partial x}}+
{vA_y\frac{\partial u}{\partial x}}+
{wA_z\frac{\partial u}{\partial x}}-\xi \frac{A_y v^2}{x V_f}
=\nonumber \\
-\frac{1}{\rho}\frac{\partial p}{\partial x}+
G_x+f_x-b_x-\frac{R_{sor}}{\rho V_f}(u-u_w-\delta \cdot u_s)
\label{eq:2a}
\end{eqnarray}
\begin{eqnarray}
\frac{\partial u}{\partial t}+\frac{1}{V_F}
{uA_x\frac{\partial u}{\partial x}}+
{vA_y\frac{\partial u}{\partial x}}+
{wA_z\frac{\partial u}{\partial x}}+\xi \frac{A_y uv}{x V_f} = \nonumber \\
-\frac{1}{\rho}R\frac{\partial p}{\partial y}+
G_y+f_y-b_y-\frac{R_{sor}}{\rho V_f}(v-v_w-\delta \cdot v_s)
\label{eq:2b}
\end{eqnarray}
\begin{eqnarray}
\frac{\partial u}{\partial t}+\frac{1}{V_F}
{uA_x\frac{\partial u}{\partial x}}+
{vA_y\frac{\partial u}{\partial x}}+
{wA_z\frac{\partial u}{\partial x}}= 
-\frac{1}{\rho}\frac{\partial p}{\partial z}+ \nonumber \\
G_z+f_z-b_z-\frac{R_{sor}}{\rho V_f}(w-w_w-\delta \cdot w_s),
\label{eq:2c}
\end{eqnarray}
where, $(G_x, G_y, G_z)$ are so called body accelerations
\cite{flow3d},  $(f_x, f_y, f_z)$ are
viscous accelerations,  $(b_x, b_y, b_z)$ are flow losses in porous
media or across porous baffle plates, and the final term accounts for
the injection of mass at a source represented by a geometry component.
As we mentioned above, FLOW3D is a general purpose fluid dynamics program,
which includes many specific situations. However, in this short
description we try to make a valuable impression about FLOW3D 
and give here the general form of all equations.
Next,  the term $U_w=(u_w,v_w,w_w)$ in Eqn. (5)
is the velocity of the source component, which will generally be
non-zero for a mass source at a General Moving Object (GMO) \cite{flow3d}.
The term $U_s=(u_s,v_s,w_s)$ is the velocity of the fluid at the surface 
of the source relative to the source itself. It is
computed in each control volume as 
\begin{equation}
\vec U_s=\frac{1}{\rho_s}\frac{d(Q \vec n)}{dA}
\end{equation}
where $dQ$ is the mass flow rate, $\rho_s$ fluid source density, 
$dA$ the area of the source surface in the cell
and $\vec n$ the outward normal to the surface. 

The source is of the stagnation pressure type
when in Eqs. (\ref{eq:2a}-\ref{eq:2c}) $\delta=0.0$. Next,
$\delta=1.0$ corresponds to the source of the static pressure type.

It is assumed, that at a stagnation pressure source fluid
enters the domain at zero velocity. As a result, pressure
should be considered at the source to move the fluid away from 
the source. For example, such sources are designed to model fluid
emerging at the end of a rocket or the simple deflating process of a balloon. 
In general, stagnation pressure sources apply to cases
when the momentum of the emerging fluid is created 
inside the source component, like in a rocket engine.
At a static pressure source the fluid velocity is 
computed from the mass flow rate and the surface area of the
source. In this case, no extra pressure is required 
to propel the fluid away from the source. An example of
such a source is fluid emerging from a long straight pipe. 
Note that in this case the fluid momentum is created
far from where the source is located.

For a variable dynamic viscosity $\mu$, the viscous accelerations are
\begin{eqnarray}
\rho V_F f_x = w^s_x - (
\frac{\partial}{\partial x}(A_x\tau_{xx}) +
R\frac{\partial}{\partial y}(A_y\tau_{xy}) + \nonumber \\
\frac{\partial}{\partial z}(A_z\tau_{xz}) + 
\frac{\mu}{x}(A_x\tau_{xx}-A_y\tau_{yy})
)
\end{eqnarray}
\begin{eqnarray}
\rho V_F f_y = w^s_y - (   
\frac{\partial}{\partial x}(A_x\tau_{xy}) +
R\frac{\partial}{\partial y}(A_y\tau_{yy}) + \nonumber \\
\frac{\partial}{\partial z}(A_z\tau_{yz}) + 
\frac{\mu}{x}(A_x+A_y\tau_{xy})
)
\end{eqnarray}
\begin{eqnarray}
\rho V_F f_z = w^s_z - \left (
\frac{\partial}{\partial x}(A_x\tau_{xz})+
R\frac{\partial}{\partial y}(A_y\tau_{yz})\right ) + \nonumber \\
\left (\frac{\partial}{\partial z}(A_z\tau_{zz}) + 
\frac{\mu}{x}(A_x\tau_{xz})
\right ),
\end{eqnarray}
where
\begin{eqnarray}
\tau_{xx}=-2\mu\left(\frac{\partial u}{\partial
x}-\frac{1}{3}\left(\frac{\partial u}{\partial x}+
R\frac{\partial v}{\partial y}+
\frac{\partial w}{\partial z}+
\frac{\xi u}{x}\right)\right)
\label{eq:wss1}
\end{eqnarray}
\begin{eqnarray}
\tau_{yy}=-2\mu\left(\frac{\partial v}{\partial
x}+\xi\frac{u}{x}\right ) \nonumber \\
-\frac{1}{3}\left(\frac{\partial u}{\partial x}+
R\frac{\partial v}{\partial y}+
\frac{\partial w}{\partial z}+
\frac{\xi u}{x}\right )
\end{eqnarray}
\begin{eqnarray}
\tau_{zz}=-2\mu\left(\frac{\partial w}{\partial
z}-\frac{1}{3}\left(\frac{\partial u}{\partial x}+
R\frac{\partial v}{\partial y}+
\frac{\partial w}{\partial z}+
\frac{\xi u}{x}\right)\right)
\end{eqnarray}
\begin{eqnarray}
\tau_{xy}=-\mu\left(\frac{\partial v}{\partial x}+
R\frac{\partial u}{\partial y}-\frac{\xi v}{x}\right)
\end{eqnarray}
\begin{eqnarray}
\tau_{xz}=-\mu\left(\frac{\partial u}{\partial z}+
\frac{\partial w}{\partial x}\right)
\end{eqnarray}
\begin{eqnarray}
\tau_{yz}=-\mu\left(\frac{\partial v}{\partial z}+
R\frac{\partial w}{\partial y}\right).
\label{eq:wss6}
\end{eqnarray}
In Eqs. (\ref{eq:wss1})-(\ref{eq:wss6})
the terms $w^s_x, w^s_y$ and $w^s_z$ are wall shear stresses.
If these terms are equal to zero, there is no wall shear stress. This is
because the remaining terms contain the
fractional flow areas $(A_x, A_y, A_z)$ which vanish
at the walls. The wall stresses are modeled by assuming a zero 
tangential velocity on the portion of any area
closed to flow. Mesh and moving obstacle boundaries are an 
exception because they can be assigned non-zero
tangential velocities. In this case the allowed boundary motion 
corresponds to a rigid body translation of the
boundary parallel to its surface. 
For turbulent flows, a 
law-of-the-wall velocity profile is assumed near the
wall, which modifies the wall shear stress magnitude.

As we already mentioned, in all simulations we consider the blood flow
as pulsatile flow.
The final result for the inflow waveform has been taken from work \cite{greece2007}.
The waveform is shown in Fig. 1. These velocity values are used  as
time-dependent inflow initial boundary conditions. These numbers are included
directly in the FLOW3D program.

\begin{figure}[ht]
\begin{center}
\includegraphics*[scale=1.0,width=18pc,height=14pc]{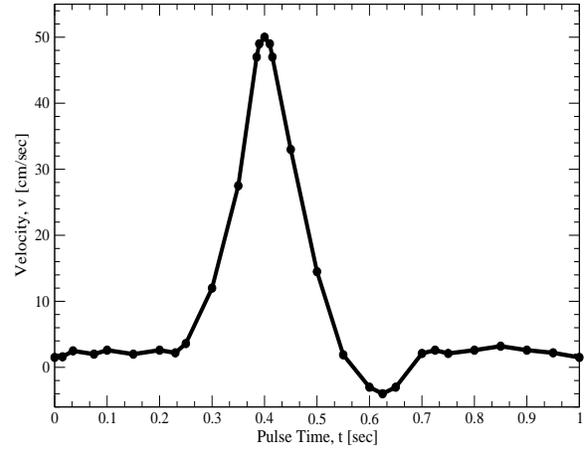}
\caption{Velocity waveform at the vessel inlet. Results taken from Fig. 2 of
work \cite{greece2007}.}
\end{center}
\label{fig:fig1}
\end{figure}

The equations of fluid dynamics should be solved together with
specific boundary conditions. 
%
The numerical model starts with a computational mesh, or grid. It consists of a 
number of interconnected elements, or 3D-cells.
These 3D-cells subdivide the physical space into 
small volumes with several nodes associated 
with each such volume. The nodes are used to store values of the 
unknown parameters, such as pressure, strain rate,
temperature, velocity components etcetera. This procedure provides values for defining the flow
parameters at discrete locations and to set up specific boundary
conditions. Finally, one can start developing effective numerical 
approximations for the solution  of the fluid dynamics equations. 

New pressure-velocity solvers have been implemented in FLOW-3D.
We used the so called GMERS method.   
GMRES stands for the
generalized minimum residual method. In addition to the GMRES solver, 
a new optional algorithm, the
generalized conjugate gradient (GCG) algorithm, has also been 
implemented for solving viscous terms in the
new GMRES solver. This new solver is a highly accurate and 
efficient method for a wide range of problems.
It possesses good convergence, symmetry and speed properties; 
however, it does use more memory than the
SOR or SADI methods. The GMRES solver does not use any over- or
under-relaxation \cite{flow3d}.

\Section{Results}

This section represents the results of our simulations for blood
flows in a simple vessel and in the human aortic arch. The first geometry was
chosen for preliminary test calculations and testing the FLOW3D
program. One of the most important preliminary testing tasks
is the check numerical convergence. This test has been
successful and our results will be shown below in this paper.
To our best knowledge the current work is a first attempt to 
apply FLOW3D to human blood flows in vessels and in aortic arch.

First, we present results for
a simpler geometry vessel in the shape of a tube. However, the human
blood is treated as real and a non-Newtonian liquid. The necessary data
for viscosity of the blood
we found from previous laboratory and clinical measurements \cite{1991,9}.
We take into account the real pulsatile flow,
which is shown in Fig. 1. The data for Fig. 1 have also been obtained in 
clinical measurements \cite{greece2007}.

After such preliminary simulations we
switch to a more complicated spatial configuration. In this work it is
the aortic
arch. It is axiomatic that real people may have different
size aortic arches with slightly different shapes. However,
we carried out simulations for an average size and shape aortic arch.

The main goal of this work is to treat the above mentioned systems
realistically, reveal the physics of the blood flow dynamics,
and to obtain reliable results for pressure, dynamic
viscosity, velocity profiles and strain rate distributions. 
Also, we tested, the widely cited in literature, Newtonian and non-Newtonian models
of the human blood. 

\subsection{Blood flow in vessel}

As we mentioned above
in this work we adopted the shape of a straight vessel as a tube. The sizes
of the tube are: $L=8$ cm in length and $R=0.34$ cm in the inner radius. The thickness
of the vessel wall is $s=0.03$ cm. We have chosen 5.5 cycles of the blood pulse.

Consider in more detail the expression (\ref{eq:mu1}). In these calculations
we follow the works \cite{1991,9}, where the Carreau model of the
human blood has also been used. In consistence with \cite{1991,9} we choose
the following coefficients:
$\lambda_2=\lambda_{00}=0$, $a=0$ and $E_T=1$, that is we don't take into
account the temperature dependence of the viscosity. Next: $\lambda_0=1$,
$\lambda_1=3.313$ sec, $\mu_{\infty}=0.0345$ P, $\mu_0=0.56$ P, and $n=0.3568$.

In our calculations
we applied a cylindrical coordinate system with the axis $OZ$ directed over
the tube axis. Different numbers of cells have been used to discretize the empty
space inside the tube. In the open space (inner part of
the tube) the fluid dynamics equations have been solved together with appropriate
mathematical boundary conditions. The convergence was achieved when
we used 52,800 cells, that is we used 100 points over $OZ$, 22 points
over the radius of the inside space $R=0.34$ cm, and 24 points over
azimuthal angle $\Phi$ from 0 to 2$\pi$.

Time-dependent results for pressure, strain rate and 
velocity component $W$ are presented in
Fig. 2. We chose three precise geometrical points to compare results - 1st: in the
inlet of the tube, when $Z=0$; 2nd: in the middle point $Z=4.0$ cm, and
3rd: in outlet point $z=8.0$ cm. The data for Fig. 2 was obtained with two different
models of human blood. The bold lines are results with the non-Newtonian viscosity
(\ref{eq:mu1}) and the dashed lines are results with the Newtonian model
when the viscosity $\mu$ has a constant value and equals to 0.0345 P.
As one can see the results are different for
strain rate distributions and very different for pressure
distributions. These results clearly indicate that probably in most
cases of computer simulations of human blood flows only the non-Newtonian
model should be used.

In Fig. 3 we show our time-dependent convergence results. These
data are obtained with the realistic non-Newtonian model of the blood.
Again, we have chosen three geometrical points to compare results: 1- in the
inlet of the tube, when $Z=0$; 2- in the middle point $Z=4.0$ cm, and
3- outlet point $z=8.0$ cm.

Three upper rows represent time-dependent plots for
pressure, dynamics viscosity and strain rate distributed over the
simulation time, which is 5.5 sec. However, below in
three lower rows results for the kinematic characteristics of the blood
flow are also shown, which are three components of the blood velocity:
$U, V, W$. The results are obtained with three sets of computation
cell distributions, where we used 36,000 cubic cells, 47,000 and
52,000 cubic cells.
As we see from Fig. 3 it is harder to obtain convergence for the pressure
distributions and easier for other shown parameters.

\begin{figure}[ht]
\includegraphics*[scale=1.0,width=19.5pc,height=22pc]{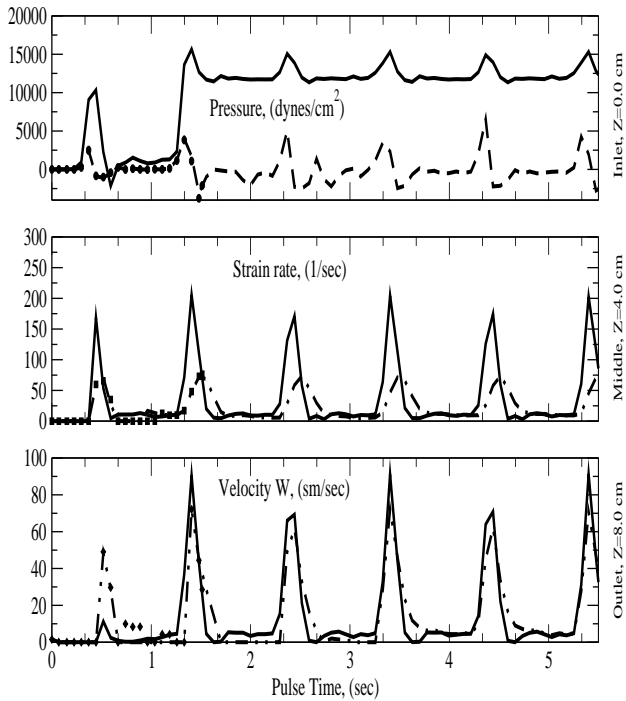}
\caption{Time-dependent pressure, strain rate and velocity
component W - over OZ axis. Results for straight vessel for 3 spatial
points: Z=0, 4, 8 cm.
Bold lines: calculations with realistic non-Newtonian viscosity of
human blood. Dashed lines: simple Newtonian approximation.}

\label{fig:fig2}
\end{figure}
\begin{figure}[ht]
\includegraphics*[scale=1.0,width=37pc,height=28pc]{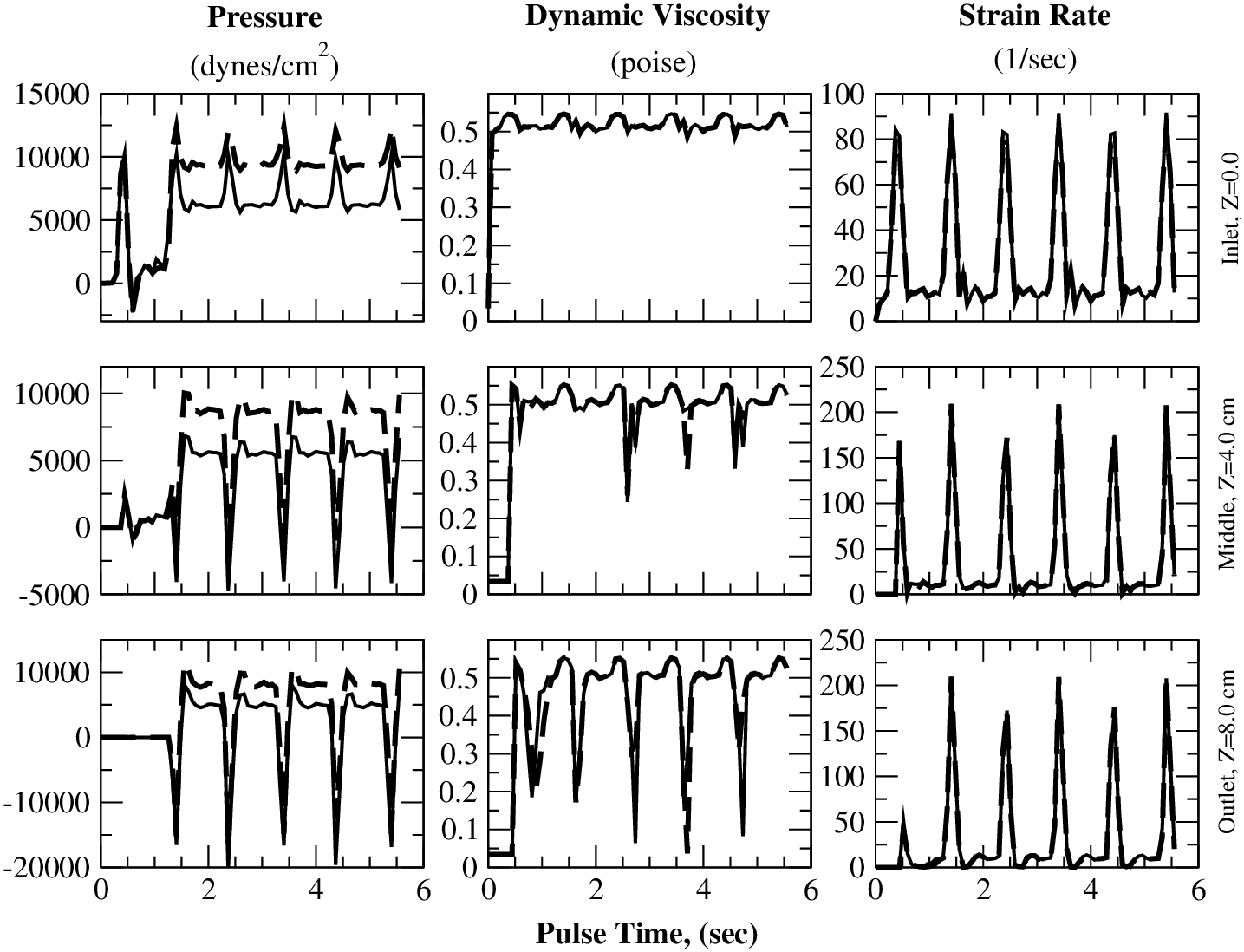}
\vspace{3mm}\\
\includegraphics*[scale=1.0,width=37pc,height=28pc]{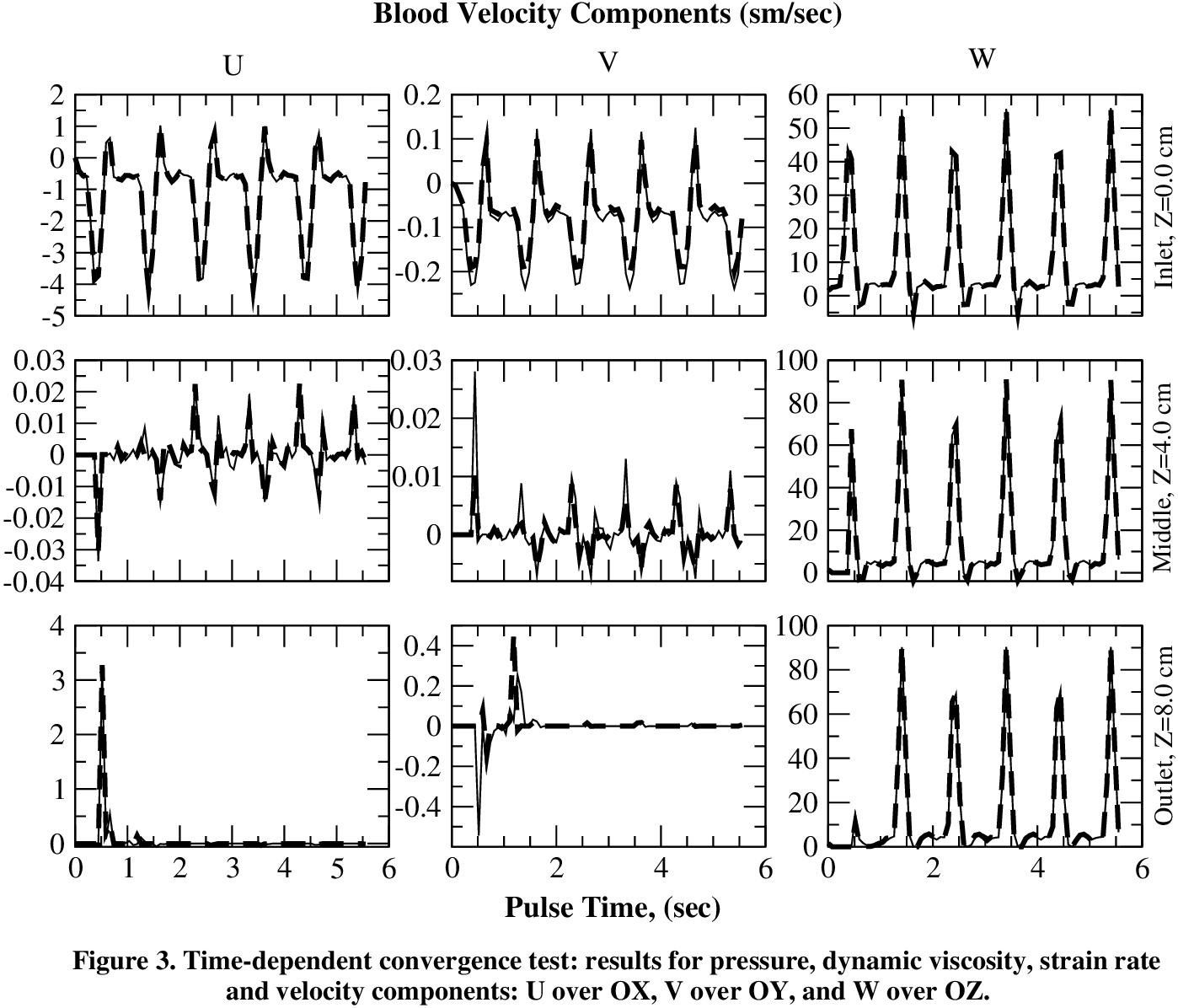}
\label{fig:fig3}
\end{figure}

\subsection{Blood flow in aortic arch}

The aortic arch is represented as a curved tube \cite{morris2005}. 
In our simulations the outer radius of the tube is 2.6 cm. A straight vessel (tube) is
also merged to the arch. The length of the straight tube is about 4 cm. Again,
the thickness of the wall is 0.03 cm, and the inner radius of the tube
is $r=0.34$ cm.
The geometry is shown in Figs. 4 and 5. This configuration closely models and
represents the real aortic arch. One of the goals of these simulations is to
reveal the physics of the blood flow dynamics in the arch.

Now we use the Cartesian coordinate system. Here we also
carried out a convergence test. To better represent the shape of the arch
we applied five Cartesian sub-coordinate systems in our FLOW3D simulations.
After the discretization the total number of all cubic cells reached about 400,000.
It is important to mention here, that we again obtained a full
numerical convergence.

As we mentioned the goal of these simulations is to compute pressure,
velocity and strain rate distributions in the arch, while the human
blood is treated as a non-Newtonian liquid and the realistic pulsatile
blood flow is used as it is shown in Fig. 1.
In Figs. 4 and 5 we show the results for strain rate distributions inside the arch
for four specific time moments. At the
most left point, which is inlet, we specify the pulsatile velocity
source as the initial condition, that is the data from Fig. 1 are used. From
the general theory of fluid mechanics \cite{landau} it is possible to determine
together with viscosity and spatial geometry,
the dynamics of the blood according to the Navier-Stokes equation and its
boundary conditions.
Small vectors indicate the blood velocity. As can be seen blood flows
from left to right in direction. However, because of pulsatility
blood flows in the opposite direction too. It is seen in
Fig. 4 - lower right graph \# 43. It is in good agreement
with the general physical intuition, and it additionally shows
correctness of these simulations. 

The values of the strain
rate are also shown. These values are strongly oscillating. From
the plots one can conclude that in the region of the arch the strain
rate values are becoming much larger than in the region of the straight vessel.
This result represents clear evidence that in this
part of the human vascular system atherosclerotic plaques should localize less than in
the straight vessels.
However, the higher wall shear stress values in the aortic arch could be
the reason for sudden mechanical disruption of the arterial wall in this
part of the human vascular system. These results are consistent
with laboratory and clinical observations.

In conclusion, we would like to point out here, that the developments in
this work can be directly applied to even more interesting and
important situation such as, when a stent is implanted inside
a vessel \cite{Duraiswamy2007,10a,11a}.
In that case, for example, it is important to determine
blood flow disturbance, the pressure distribution, strain rate and etcetera.
This work is in progress in our group.

\clearpage
\begin{figure}[ht]
\vspace{-3.3cm}
\begin{center}
\includegraphics[scale=1.0,width=20pc,height=25pc]{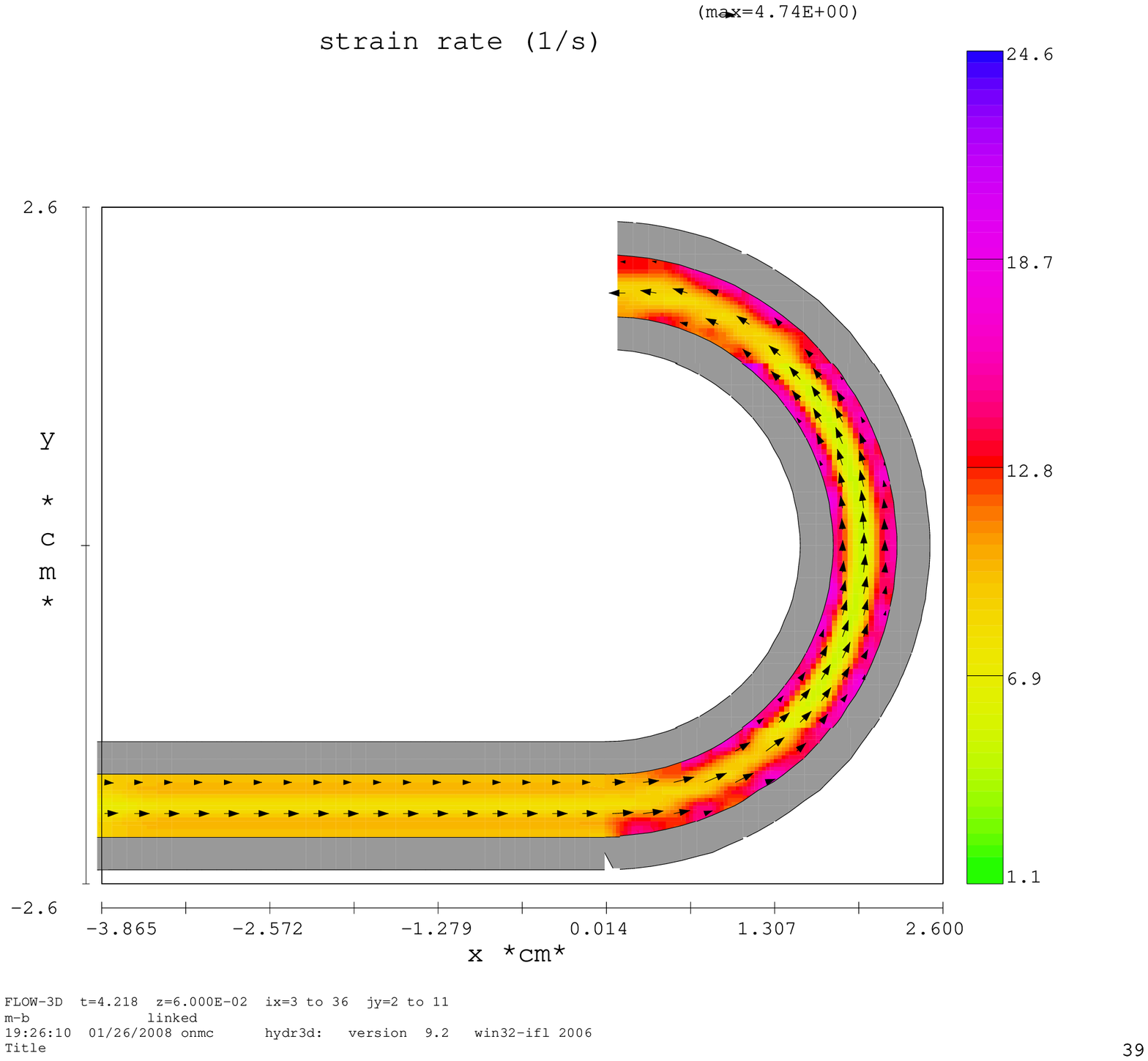}
\includegraphics[scale=1.0,width=20pc,height=25pc]{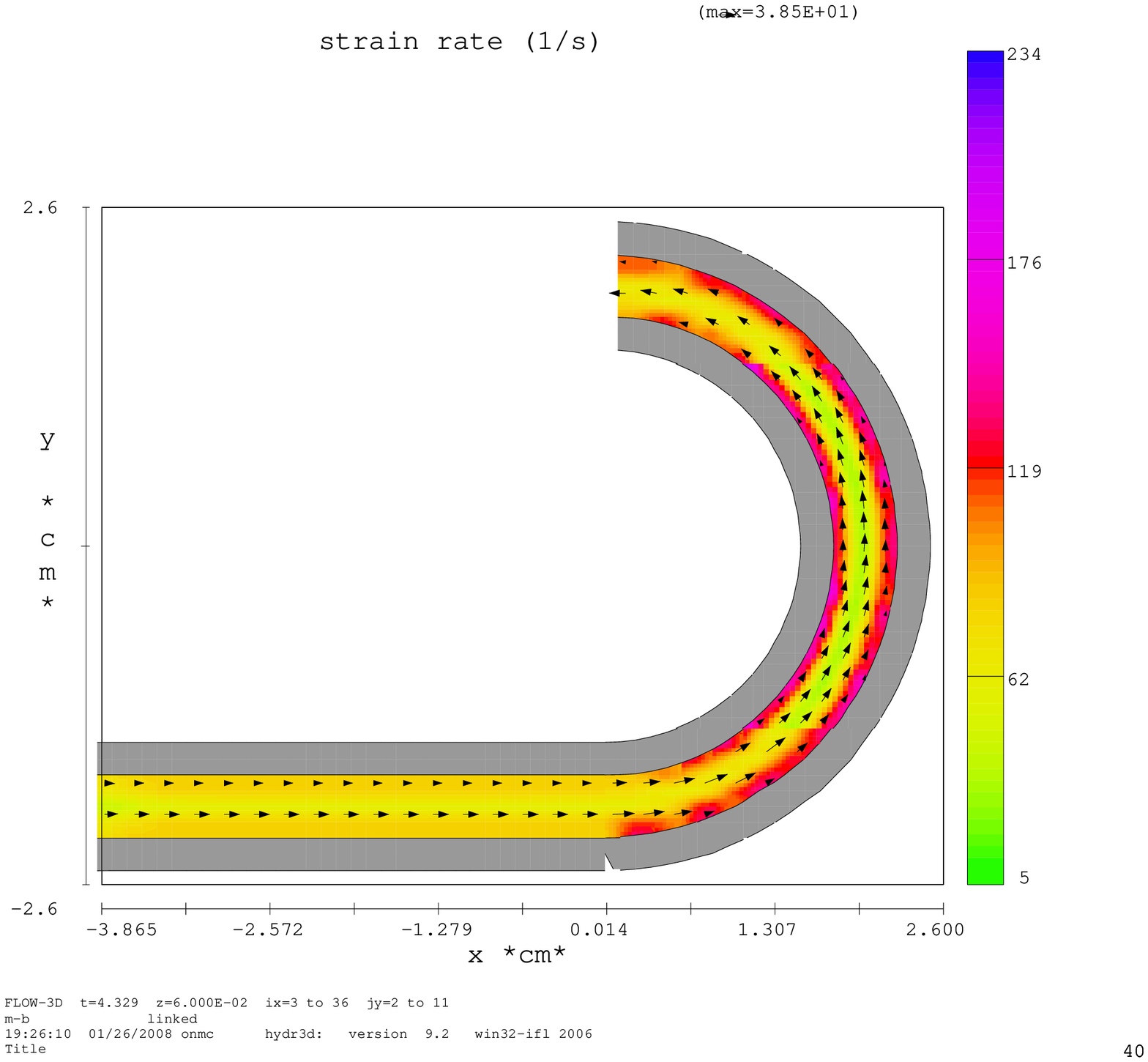}\\
\end{center}
{\bf Figure 4. Blood flow in the aortic arch for two consecutive moments of the
discretized time. A strong pulsatility of the strain rate values is seen:
Upper plot shows results for t = 4.218 sec, where strain rate ranges
from e = 1.1 1/sec to e = 24.6 1/sec. Lower plot shows results for t =
4.329 sec, where
strain rate ranges from e = 5. 1/sec to e=234. 1/sec.
The maximum values of the strain rate are localized in the region inside
the arch. Blood flows from right to left in both pictures.}
\label{fig:fig4}
\end{figure}
\begin{figure}[ht]
\vspace{-3.3cm}
\begin{center}
\includegraphics[scale=1.0,width=20pc,height=25pc]{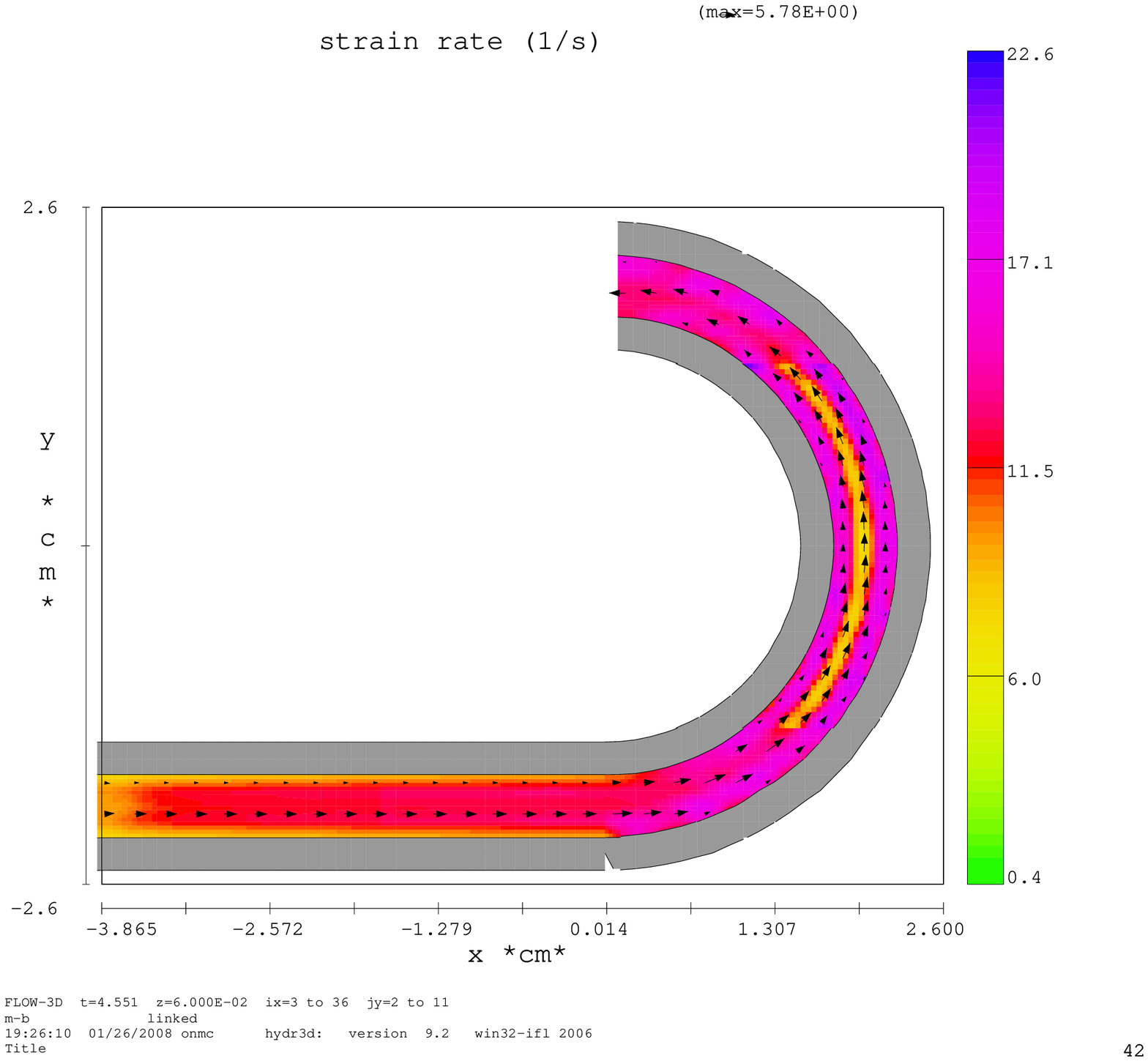}
\includegraphics[scale=1.0,width=20pc,height=25pc]{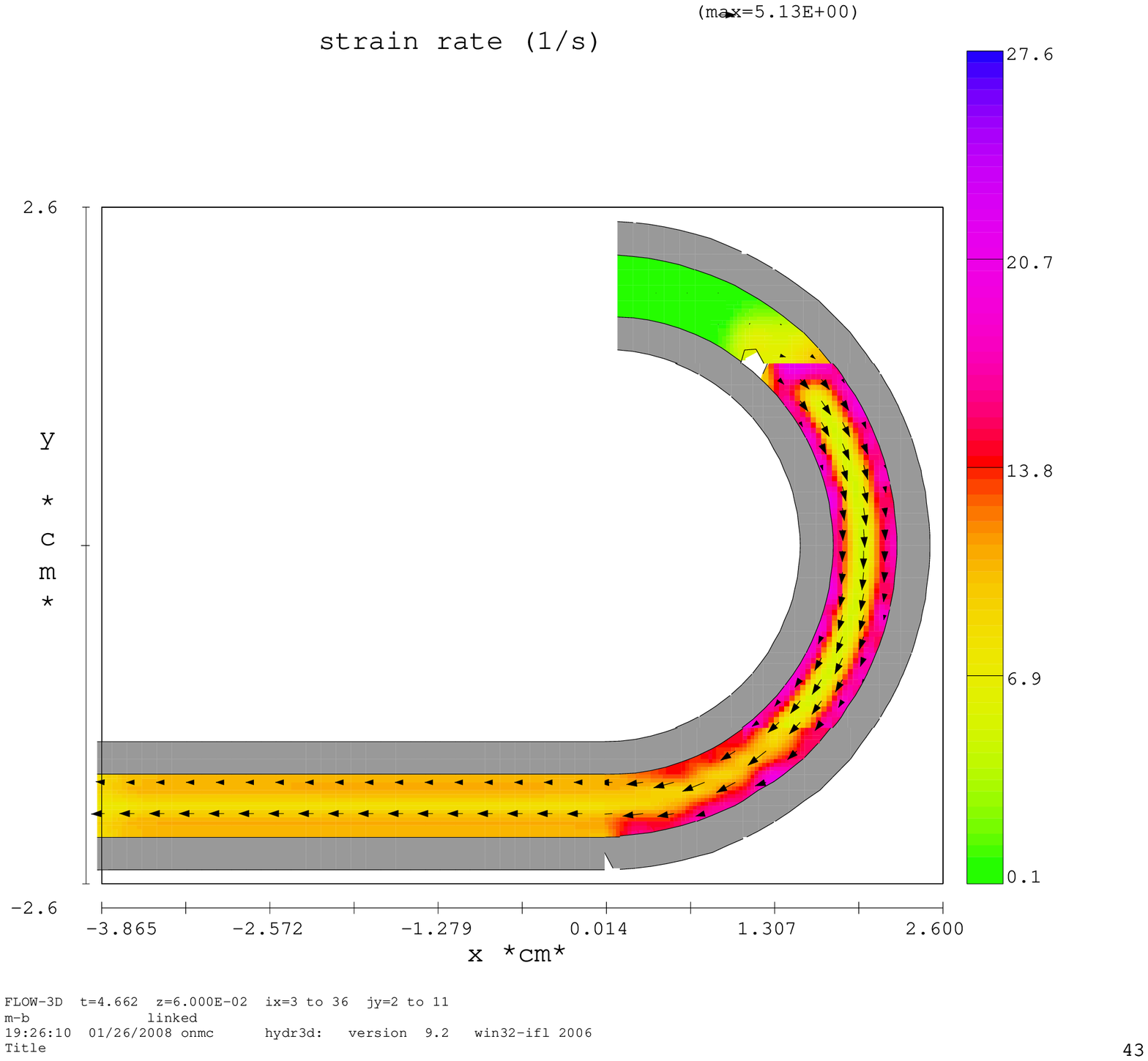}\\
\end{center}
{\bf Figure 5. Blood flow in the aortic arch for two consecutive moments of
discretized time. 
Upper plot shows results for t = 4.551 sec, where strain rate ranges
from e = 0.4 1/sec to e = 22.6 1/sec. Lower plot shows results for t =
4.662 sec, where
strain rate ranges from e = 0.1 1/sec to e = 27.6 1/sec.
The maximum values of the strain rate are localized again in the region inside
the arch. In the upper plot blood flows from left to right, however in
the lower plot: from right to left.}
\label{fig:fig5}
\end{figure}



\begin{thebibliography}{999}
\bibliographystyle{latex8}
\bibliography{latex8}


\bibitem{1991} Y.I. Cho, K.R. Kensey, Effects of the non-Newtonian
viscosity of blood on flows in a diseased arterial vessel. Part 1:
Steady Flows, Biorheology 28, 241-262 (1991).



\bibitem{19b}
Taylor, C.A., Draney, M.T.,
Experimental and Computational Methods in Cardiovascular Fluid Mechanics,
Annual Review of Fluid Mechanics 36, 197-231 (2004).

\bibitem{9}
Johnson, B.M., Johnson, P.R., Corney, S., and Kilpatrick, D.,
Non-Newtonian Blood Flow in Human Right Coronary Arteries: Steady State
Simulations, J. of Biomech. 37, 709 (2004).

\bibitem{morris2005} 
Morris, L., Delassus, P., Callan, A., Walsh, M., Wallis, F., Grace,
P., McGloughlin, T., 3-D Numerical Simulation of Blood Flow Through
Models of the Human Aorta, J. Biomechanical Engineering 127, 767-775 (2005).



\bibitem{Duraiswamy2007}
Duraiswamy, N., Schoephoerster, R.T., Moreno, M.R., Moore Jr., J.E.,
Stented Artery Flow Patterns and Their Effects on the Artery Wall,
Annual Review of Fluid Mechanics 39, 357-382 (2007).

\bibitem{greece2007}
Y. Papaharilaou, J.A. Ekaterinaris, E. Manousaki, A.N. Katsamouris,
A Decoupled Fluid Structure Approach for Estimating Wall Stress in Abdominal
Aortic Aneurysms, J. Biomechanics 40, 367-377 (2007).



\bibitem{10a}
Faik, I., Mongrain, R., Leask, R.L., Rodes-Cabau, J., Larose, E.,
Bertrand, O.,
Time-Dependent 3D Simulations of the Hemodynamics in a Stented Coronary
Artery, Biomedical Materials 2 (1), art. no. S05, S28-S37 (2007).  

\bibitem{11a}
Banerjee, R.K., Devarakonda, S.B., Rajamohan, D., Back, L.H.,
Developed Pulsatile Flow in a Deployed Coronary Stent,
Biorheology 44 (2), 91-102 (2007).                    




\bibitem{flow3d}
FLOW-3D Users Manual, Version 9.2, Flow Science, Santa Fe, New Mexico, 2007.

\bibitem{landau}
Landau, L.D., Lifshitz, E.M., Fluid Mechanics, Volume 6, Pergamon
Press Ltd., 1959.




\end{thebibliography}
\end{document}